\documentclass[letter]{aa}

\usepackage{txfonts}
\usepackage{epsf}
\usepackage{graphicx}
\usepackage{color}

\def \etal   {\hbox{et~al.\/}}

\newcommand{\blyr}{\mbox{$\beta\,$Lyr}}

\begin{document}

\title{Phase-dependent X-ray observations of the $\beta$ Lyrae system:}
\subtitle{No eclipse in the soft band}

\author{R.~Ignace\inst{1}
\and L.~M.~Oskinova\inst{2}
\and W.~L.~Waldron\inst{3}
\and J.~L.~Hoffman\thanks{NSF Astronomy and Astrophysics
Postdoctoral Fellow}\inst{3,4}\thanks{Currently at Department of Physics
and Astronomy, University of Denver}
\and W.-R.~Hamann\inst{2}}

\institute{Department of Physics, Astronomy, \& Geology,
East Tennessee State University, Johnson City, TN 37614
\and 
Lehrstuhl Astrophysik der Universit\"at Potsdam,
Am Neuen Palais 10, D-14469 Potsdam, Germany
\and Eureka Scientific, Inc., 2452
Delmer Street Suite 100, Oakland, CA 94602-3017
\and Department of Astronomy, UC Berkeley, 601 Campbell
Hall, Berkeley, CA 94530
}

\offprints{R.~Ignace, \email{ignace@etsu.edu}}

\date{Received / Accepted }

\abstract
{}
{We report on observations of the eclipsing and interacting
binary $\beta$ Lyrae from the {{\it Suzaku}} X-ray telescope.  
This system involves an early B star
embedded in an optically and geometrically
thick disk that is siphoning atmospheric gases from a less massive
late B~II companion.}
{Motivated by an
unpublished X-ray spectrum from the {{\it Einstein}} X-ray telescope
suggesting unusually hard
emission, we obtained time with Suzaku for pointings at three
different phases within a single orbit. }
{From the XIS detectors, the softer X-ray emission appears typical
of an early-type star.  What is surprising is the remarkably 
unchanging character of this emission, both in luminosity and in spectral
shape, despite the highly asymmetric geometry of the system.  We see
no eclipse effect below 10~keV.  
The constancy of the soft emission is plausibly related to the wind
of the embedded B star and Thomson scattering of X-rays in the
system, although it might be due to extended shock structures arising
near the accretion disk as a result of the unusually high mass-transfer
rate.  There is some evidence from the PIN instrument for hard emission
in the 10-60~keV range.  Follow-up observations with the RXTE satellite
will confirm this preliminary detection.}
{}


\maketitle

\section{Introduction}

The binary $\beta$ Lyr is a nearly edge-on, semi-detached interacting
system that has undergone mass reversal and remains in a phase of
large-scale mass transfer.  The primary, mass-losing star (the ``Loser'')
is a B6-B8 IIp star. The mass-gaining star (the ``Gainer'') is embedded
in an optically thick accretion disk and is not directly visible.
Although the embedded source had been considered as a possible compact
object (Devinney 1971; Wilson 1971), it is probably a main sequence B0 star
(Hubeny \& Plavec 1991).  The system is very complex, having bipolar
jet-like structures (Harmanec \etal\ 1996; Hoffman \etal\ 1998),
a circumbinary envelope (Batten \& Sahade 1973; Hack \etal\ 1975),
and a substantial kilo-Gauss magnetic field (Leone \etal\ 2003).

\begin{table}
\begin{center}
\caption{Properties of $\beta$ Lyrae \label{tab1}}
\begin{tabular}{lcc}
\hline\hline
Component$^a$ & Gainer & Loser \\
\hline
\rule[0mm]{0mm}{3.5mm}$T_{\rm eff}$[K] & 32,000 & 13,300 \\
\rule[0mm]{0mm}{4mm}Spectral Type & $\approx$B0 V & B6-8 IIp \\
\rule[0mm]{0mm}{4mm}$M/M_\odot$ & $\approx$13 & $\approx$3 \\
\rule[0mm]{0mm}{4mm}$\dot{M}^b[M_{\odot}~{\rm yr}^{-1}]$ & $4.7 \times 10^{-8}$ & $7.2 \times 10^{-7}$ \\
\rule[0mm]{0mm}{4mm}$ \varv_\infty^b$[km s$^-1$] &  1470  & 390 \\  
\\\hline\hline
System Property$^a$ & \multicolumn{2}{c}{Value} \\ \hline
\rule[0mm]{0mm}{3.5mm}Orbital Period & \multicolumn{2}{c}{12.9 days}\\
\rule[0mm]{0mm}{3.5mm}Viewing Inclination & \multicolumn{2}{c}{$86^\circ$}\\
\rule[0mm]{0mm}{4mm}Binary Separation & \multicolumn{2}{c}{$55-60 R_\odot$}\\
\rule[0mm]{0mm}{4mm}Distance & \multicolumn{2}{c}{270 pc} \\
\rule[0mm]{0mm}{4mm}$\log N_H$ (cm$^{-2}$)  & \multicolumn{2}{c}{20.76}  \\
\rule[0mm]{0mm}{4mm}{\it ROSAT} PSPC$^c$: & \multicolumn{2}{c}{0.07 cps} \\
\rule[0mm]{0mm}{4mm}{\it Einstein} SSS$^d$: & \multicolumn{2}{c}{0.11 cps}\\
\hline 
\end{tabular}

\parbox{2.5in}{\small $^a$ Component and system properties taken
from Harmanec (2002) unless otherwise noted. \\
$^b$ Mass-loss rate and terminal speed for the stellar
winds of the respective binary components (Mazzali 1987).\\
$^c$ Bergh\"{o}fer \&
Schmitt 1994)\\ 
$^d$ Waldron (private comm.)}
\end{center}
\end{table}

\begin{figure}[t]
  \resizebox{\hsize}{!}{\includegraphics{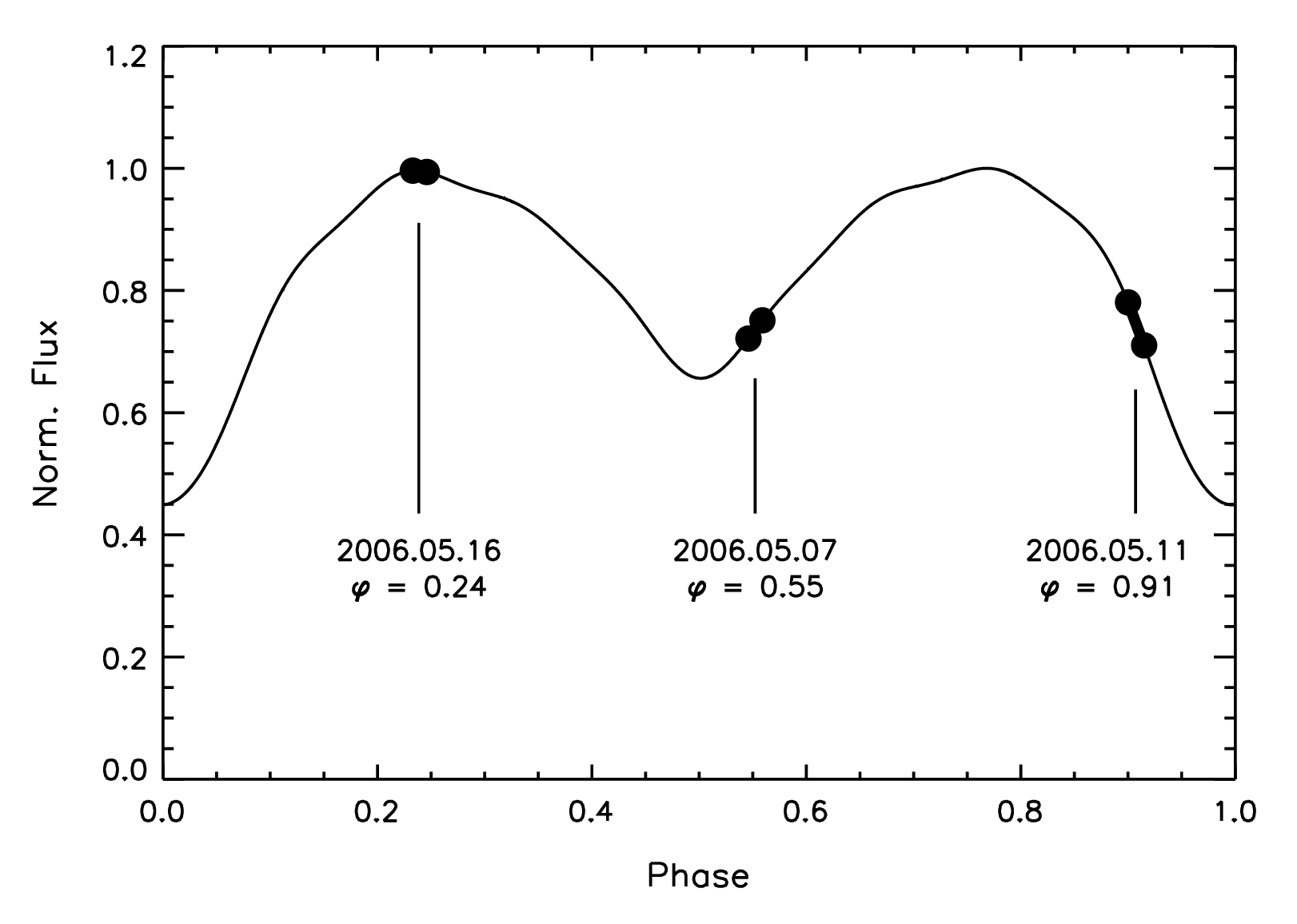}}
\caption{
Illustration of the Suzaku pointings with
respect to the binary phase.  The solid line represents the normalized
V-band light curve (Fourier fit by Harmanec et al. 1996). 
Primary eclipse (phases 0.0 and 1.0) occurs when
the mass-losing star is occulted by the disk.
The three intervals marked on the curve represent our Suzaku pointings,
which occurred during a single orbit in May 2006.  Phases were calculated
from the quadratic ephemeris of Harmanec \& Scholz (1993).
\label{fig1}}

\end{figure}

The optical light curve of the system features a primary minimum that
is $\approx 1$ magnitude deep and a secondary minimum $\approx 0.4$
magnitudes deep (see Fig.~\ref{fig1}); however, the secondary minimum
is deeper than the primary minimum at shorter wavelengths, and below
Ly$\alpha$ the eclipses no longer appear (Kondo \etal\ 1994).  A summary
of the system properties is given in Table~\ref{tab1}.  The orbital period
is 12.9~days, and the mean light curves appear stable with epoch.
The UV spectrum of $\beta$ Lyr is dominated by an anomalous continuum
and emission lines with unusually strong P~Cygni profiles typical of
hot star winds (Hack \etal\ 1975; Aydin \etal\ 1988; Mazzali 1987).

Despite numerous and ongoing modeling attempts (e.g., Wilson 1974;
Linnell \& Hubeny 1996; Bisikalo \etal\ 2000; Linnell 2002; Nazarenko \&
Glazunova 2003, 2006ab), no model is yet capable of matching the observed
light curves of $\beta$ Lyr from the IR through the UV.  Strangely,
$\beta$~Lyr has been largely unstudied in the X-ray regime, despite
the strong X-ray flux detected by the {\it ROSAT} HRI (Bergh\"{o}fer \&
Schmitt 1994).  An unpublished spectrum taken with the {\it EINSTEIN}/SSS
in 1979 reveals X-ray emission at relatively high energies, suggesting
that phase-dependent observations may provide new clues to resolving
the puzzle of the $\beta$ Lyr geometry and interactions.  Exploiting
{\it Suzaku's} excellent sensitivity to hard X-rays, we conducted three
pointed observations of $\beta$~Lyr within the same orbit.  In the
following section the observations and reduction of data are detailed.
Analyses of the spectra with phase are described in \S 3, and a discussion
of the results is presented in \S 4.

\section{Observations and Data Reduction}

\begin{figure}
\centering
  \resizebox{\hsize}{!}{\includegraphics{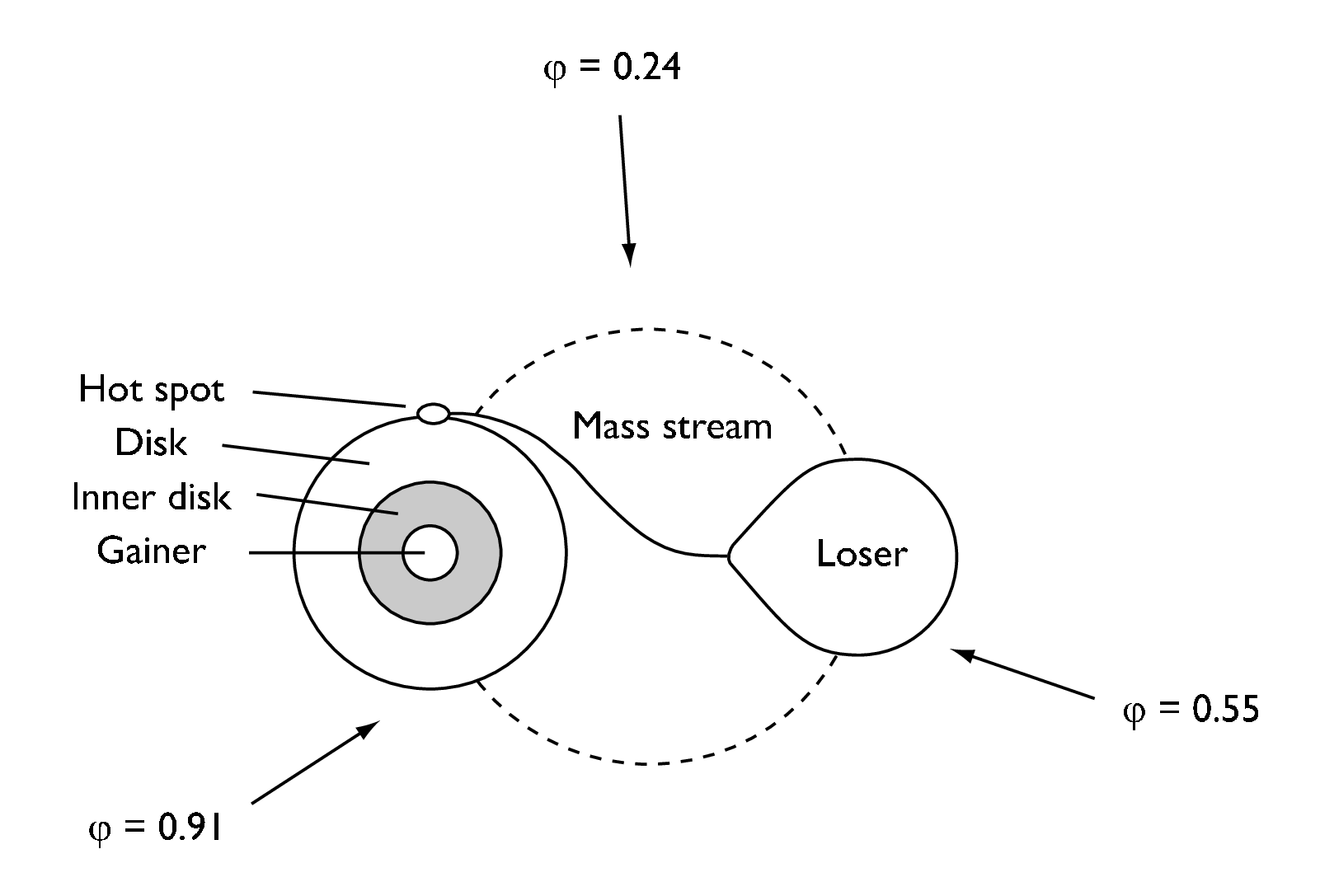}}
\caption{Graphic topview representation of the \blyr\ star-disk system,
following the model of Hoffman \etal\ (1998).  Components are illustrative
and not to scale.  The three {\it Suzaku} pointings occurred at phases
$\varphi=0.24$, 0.55, and 0.91 all within the same orbit.  Arrows indicate
the Earth's line-of-sight for these three phases.
\label{fig2}}
\end{figure}

The joint Japan/US X-ray astronomy satellite {\em Suzaku} (Mitsuda
\etal\ \cite{mit}) observed \blyr\ in May 2006 on three occasions at
orbital phases as shown in Fig.~\ref{fig1}, with corresponding viewing
perspectives of the \blyr\ system illustrated in Fig.~\ref{fig2}.
(Note that the primary minimum occurs when the loser star is eclipsed,
and the secondary minimum when the disk component is eclipsed.)
The pointings were spaced approximately 4.3~days apart to sample the
full 12.9~day orbit of the binary.  Exposure times and count rates are
tabulated in Table~\ref{tab2}.

{\em Suzaku} carries four X-ray Imaging Spectrometers (XIS; Koyama
\etal\ \cite{xis}) and a collimated Hard X-ray Detector (HXD; Takahashi
\etal\  \cite{hxd}).  The field-of-view (FOV) for the XIS detectors
is $17\arcmin\ \times\,17\arcmin$.  One of the XIS detectors (XIS1) is
back-side illuminated (BI) and the other three (XIS0, XIS2, and XIS3) are
front-side illuminated (FI). The bandpasses are $\sim$\,0.4\,--\,12\,keV
for the FI detectors and $\sim$\,0.2\,--\,12\,keV for the BI detector.
The BI CCD has higher effective area at low energies, however its
background level across the entire bandpass is higher compared to the
FI CCDs.

The angular resolution of the X-ray telescope onboard {\em Suzaku} is
$\approx 2\arcmin$.  Therefore in the XIS image, \blyr\ is not resolved
from two nearby B-type stars: HD\,174664 (\blyr\,B) and HD\,174639. While
the latter star was not detected by {\it ROSAT}, the former has a
{\it ROSAT} HRI count rate of $4\times 10^{-3}$ cps, as compared to the
\blyr\ {\it ROSAT} HRI count rate $4\times 10^{-2}$ cps.  We are confident
that the X-ray flux detected by {\em Suzaku} is at least 90\%
dominated by \blyr.

The HXD consists of two non-imaging instruments (the PIN
and GSO; see Takahashi \etal\ \cite{hxd}) with bandpasses of
$\sim$\,10\,--70\,keV (PIN) and $\sim40-600$~keV (GSO), and a FOV of
$34\arcmin\ \times\,34\arcmin$ (PIN).  Both of the HXD instruments are
background-limited. The background subtraction for the HXD is performed by
modeling the background spectrum.  Presently, the non-X-ray background
model (e.g., particle events) is known for the PIN detector with
$\sim$\,3-5\% accuracy (Kokubun \etal\ \cite{hxd-o}).  The background
modeling for GSO data is currently far less certain than for the PIN,
and so we do not report on the GSO measurements.

HXD-PIN data reduction and extraction of spectra were performed using the
latest calibration sources and background models. We account for cosmic
X-ray background (CXB) in fitting the spectral models using the procedure
suggested by the {\em Suzaku} team based on a ``typical'' CXB spectrum
(see www.astro.isas.ac.jp/suzaku).  We do not correct for contributions to
the HXD background due to the South Atlantic Anomaly (SAA; see Kokubun
\etal\ \cite{hxd-o}) because our observations of \blyr\ were performed
when the background count-rate was at its lowest.

\begin{table*}
\begin{center}
\caption{XIS Observations of $\beta$ Lyrae \label{tab2}}
\begin{tabular}[t]{ccccccc}
\hline\hline
Date & $\varphi^a$    &  Exposure &
\multicolumn{4}{c}{XIS Count rate ($10^2$\,cps)} \\ 
      &     &  (ksec) & XIS0 & XIS1 & XIS2 & XIS3 \\ 
\hline
\rule[0mm]{0mm}{3.5mm}2006 May 7& 0.55 &  15.4   & 
$6.35 \pm 0.24$ & $8.53\pm 0.36$ & $6.33 \pm 0.24$ & 
$ 5.15 \pm 0.22$ \\ 
\rule[0mm]{0mm}{4mm} 2006 May 12 & 0.91 & 17.7   & 
$5.76\pm 0.21$ & $8.14\pm 0.33$ & $5.56 \pm 0.21$ & 
$4.86 \pm 0.20$ \\ 
\rule[0mm]{0mm}{4mm} 2006 May 16 & 0.24 &  15.7   & 
$6.15\pm 0.23 $ & $9.58\pm 0.36$ & $6.23\pm 0.22$ & 
$4.93 \pm 0.22$ \\ 
\hline
\end{tabular}

\parbox{5in}{\small $^a$ The orbital phases for \blyr\ were computed at the midpoint of each
observation using the quadratic ephemeris of
Harmanec \& Scholz (1993).  Phases 0.0 and 1.0 correspond to 
primary eclipse.}
\end{center}
\end{table*}

Based on the Rosat All-Sky Survey (RASS), there are a few X-ray sources in
the $34\arcmin\ \times\,34\arcmin$ PIN's field of view (FOV); 
however, our target \blyr\ is by far the brightest. The stellar
coronal X-ray sources present in the FOV are expected to be faint in
the HXD's energy range. There is also an active galactic nucleus (AGN)
in the FOV about 15\arcmin\ away from $\beta$~Lyr.  This AGN, QSO\,B1847+3330,
is cataloged in the RASS with a count-rate of $0.05$\,cps. To estimate
its potential contribution to the $10-70$~keV energy range, we adopted a
standard AGN power-law spectrum with $\Gamma\,=\,2$ and a low interstellar
absorption column of $N_{\rm H}=5\times 10^{20}\,{\rm cm}^{-2}$. The
predicted count-rate for the {\em Suzaku} PIN is $\approx 0.006$~cps. The
observed HXD-PIN count-rates are listed
in Table~\ref{tab3}.  The contribution of QSO\,B1847+3330 is between
10--30\% of the detected flux.

\section{Analysis}

\begin{figure}
\resizebox{\hsize}{!}{\includegraphics{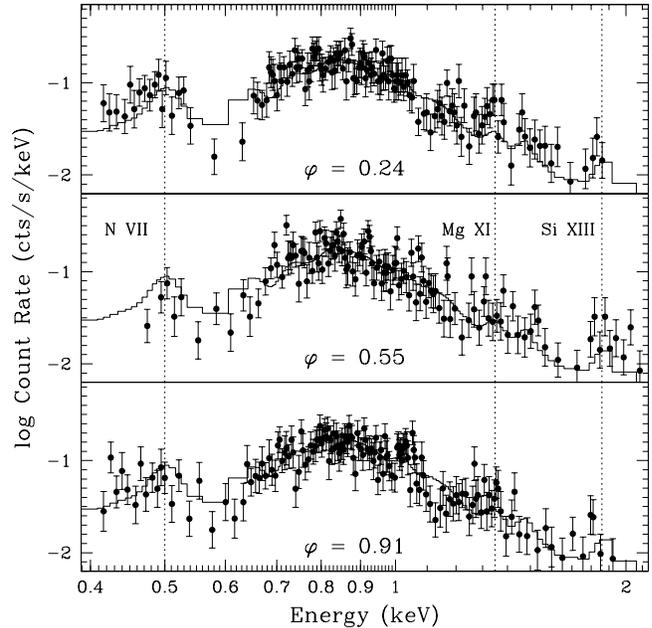}}
\caption{Spectra of $\beta$ Lyr from the XIS1 detector 
for each pointing.
Phase within the orbit is indicated in each panel.  The data are of
reasonably high signal-to-noise, and it is clear that the spectral distribution
shows little change with phase, being of nearly constant brightness and
spectral shape.	The solid line is a model fit to the data of
$\varphi=0.24$ that is replotted in the other phases for comparison.
\label{fig3}}
\end{figure}

A comparison of our three XIS spectra below 10 keV with the XIS detectors
indicates very little variability with phase (see Fig.~\ref{fig3}).
Based on chi-square model fits to the data, there is no evidence
for statistically significant differences between the three spectra.
For illustration a model fit for the $\varphi=0.24$ pointing appears
as the solid line in all three panels of Fig.~\ref{fig3}; its
good agreement with the data at all phases demonstrates that the soft
X-ray spectrum of $\beta$~Lyr is nearly constant in shape and strength.
Independent fits for each pointing are very similar.

These XIS X-ray spectra are most probably thermal in
nature, since emission lines are detected at 1.35~keV (Mg {\sc xi})
and 1.86~keV (Si {\sc xiii}).  Our fits indicate a two-component model
with the majority of the XIS X-ray emission arising from a temperature
of $\approx 7.2$~MK and a hotter but much weaker component of $\gtrsim
20$~MK.  Solar abundances were adopted except for nitrogen that was
enhanced by more than 10 times to achieve an adequate fit.  The derived
hydrogen column density from the model is $N_H = 6.5 \pm 0.24 \times
10^{20}$ cm$^{-2}$, largely consistent with the interstellar value.

Hard emission in the 10--60~keV band of the PIN is background dominated.
The observed total and the background count rates are given in
Table~\ref{tab3}.  The source counts come from the difference of the
total and background values.  Based on the total counts, there is a
significant source detection at phase $\varphi=0.55$, a non-detection
for $\varphi=0.91$, and a marginal detection at $\varphi=0.24$.
Emission at such hard energies is exceptionally unusual and unexpected
for a system like $\beta$~Lyr that consists of two early-type stars.
The implied hard X-ray luminosity above 10~keV at $\varphi\approx 0.55$
is about $10^{-3}~L_\odot$.  If confirmed, this hard component would be
an important tracer of the flow geometry and shock structure since the
circumstellar material is quite transparent to such high energy X-rays.

\begin{table}
\begin{center}
\caption{PIN Observations of $\beta$ Lyrae \label{tab3}}
\begin{tabular}[t]{cccc}
\hline\hline
$\varphi^a$    &  Exposure & Count Rate & Background \\
&  (ksec)   & (cps) & (cps) \\ 
\hline
\rule[0mm]{0mm}{3.5mm}0.55 &  14.7  & $0.623 \pm 0.007$ & $0.536 \pm 0.002$ \\
\rule[0mm]{0mm}{4mm}0.91 & 17.0  & $0.505 \pm 0.005$ & $0.509 \pm 0.002$ \\
\rule[0mm]{0mm}{4mm}0.24 &  14.1  & $0.573 \pm 0.006$ & $0.564 \pm 0.002$ \\
\hline
\end{tabular}
\end{center}
\end{table}

\section{Discussion}

The results of our {\it Suzaku} study are both revealing and perplexing.
We had expected to see an eclipse of soft X-rays and essentially no
hard emission.  Instead, we found that the soft X-rays were nearly
constant, and there is an indication of a quite hard emission component
in the system.  The near constancy of the soft spectrum suggests that
the X-ray source must be axially symmetric.  This is quite surprising,
since the system is intrinsically non-axisymmetric.

It is worth recalling that the UV lightcurve below Ly$\alpha$ is
notable for lacking any eclipses as well (Kondo \etal\ 1994).  A somewhat
analogous source to $\beta$~Lyr is the near edge-on interacting binary
W~Ser.  Weiland \etal\ (1995) interpret UV spectra of W~Ser in terms
of an extended boundary layer between the star and the accretion disk.
Thus, it may be possible that the largely steady FUV and soft X-ray
emission could arise from a similar region in the $\beta$~Lyr system.
However, we do not currently favor this explanation since the most
recent modeling of the disk and star components of \blyr\ (Linnell 2002)
indicate that gainer star is entirely obscured at all phases.

An alternative view is suggested by hydrodynamic simulations.  Bisikalo
\etal\ (2000) modeled the mass transfer of $\beta$~Lyr and found that
they could reproduce a disk-like structure but without a well-identified
``hot spot''.  Instead, their simulations yielded a complex distribution
of extended shocks.  They also were able to generate a bipolar flow
similar to the jet detected by Harmanec \etal\ (1996) and Hoffman \etal\
(1998).  However, radiative cooling was not treated self-consistently
in those simulations, but approximated through the equation of state.
It is unclear whether this model can generate gas at sufficiently high
temperatures to produce the observed X-rays.  Hydrodynamic modeling
of the $\beta$~Lyr system was also pursued by Nazarenko \& Glazunova
(2003; 2006ab), first in 2D simulations and then in 3D simulations.
These researchers included cooling curves in their simulations and
allowed for a stellar wind by the gainer.  Their models also lead to
a disk, a bipolar flow, and an environment permeated with elongated
shocks at different azimuthal orientations about the gainer.  ``Warm''
gas temperatures up to about 200,000~K are achieved, but such gas can not
contribute significant emission to the X-ray band.  The gas dynamical
simulations are qualitatively promising in terms of predicting shocked
structures that are spatially distributed in radius and azimuth throughout
the disk.  In such a model, the hot gas could potentially that hot gas
could be viewable at every phase.  However, X-ray emissions were not the
focus of those studies, and it would be useful to have new simulations
that emphasize the hot plasma structures.

We prefer an interpretation in which the X-rays arise from distributed
shocks in the wind of the gainer star.  At $kT \approx 0.6$~keV, the
spectral characteristics of the {\it Suzaku} spectra are compatible with
a typical early main sequence B star, and the XIS luminosity $L_X({\rm
XIS}) \approx 6.6\times 10^{30}$ erg s$^{-1}$ is commensurate with the
soft emission expected from an early B0-B1 star (Cohen, Cassinelli, \&
MacFarlane 1997).  Although the favored model for the binary geometry may
preclude a direct view of the gainer star, it does permit a reasonably
deep view into its wind, even at secondary eclipse (with the giant star at
front).  Moreover, we know from the polarimetric study of Hoffman \etal\
(1998) that there is substantial scattering opacity above and below the
disk plane.  The near constancy of the X-rays seen with the XIS may result
from an extended ``halo'' of scattered X-ray light, similar in spirit
to the scattered optical light observed in images of some Herbig-Haro
objects seen edge-on to their disks, such as HH~30 (Burrows \etal\ 1996).
In this picture the soft X-rays ultimately originate in the shocked wind
of the early gainer star.  Part of the emission is observed directly
from hot plasma at large radii in the wind, and part is scattered into
our line-of-sight from above and below the disk.  Thus the 0.6~keV
temperature of the hot gas would represent an upper limit value owing to
the fact that photoabsorption of X-rays is more severe toward the soft
end of the spectrum.  On the other hand, the observed X-ray luminosity
must be a lower limit. \\

We have considered three distinct origins for the observed X-ray emissions.
At this point we favor the gainer wind as the source of the soft X-rays,
but acknowledge that future hydrodynamic modeling may change this picture.
In addition, there seems to be an unusually hard component of emission in
the system, in excess of 10~keV.  Such hard emission is more typically
associated with compact objects, and not the winds of hot stars (e.g.,
Bergh\"{o}fer \etal\ 1997).  However, the nature of the gainer is still
ambiguous; perhaps the X-ray properties of \blyr\ could be understood
in relation to a central compact object (Devinney 1971; Wilson 1971).
Although this seems unlikely, the X-ray observations certainly do not
conform to our original expectations, and so at least a reconsideration of
the possibility seems in order.  The X-ray pointings by {\it Suzaku} offer
the tantalizing promise of providing new and valuable information about
the \blyr\ system, but it is clear that a far more rigorous sampling
of the X-ray lightcurve and source spectrum and new models for the
system will be needed to test the suppositions that we have put forth.
We have obtained time with the RXTE satellite to create a more complete
X-ray light curve of $\beta$~Lyr, which should provide new insight into
this complex system in the near future.

\begin{acknowledgements}

We wish to thank the referee Ed Devinney, an anonymous referee, and
Steve Shore for several valuable comments.  We are grateful for advice
given by Hugh Hudson in relation to the background of hard X-rays and
by Georg Lamer in relation to X-ray spectra of QSOs.  We also thank
Koji Mukai and Nick White for technical assistance regarding Suzaku.
RI received support from NASA grant award NNX06AI04G.  LMO acknowledges
support by the Deutsche Forschungsgemeinschaft with grant Fe\,573/3.
WLW was supported in part by NASA contract NNG07EF47P.  JLH acknowledges
the support of a NSF Astronomy \& Astrophysics Postdoctoral Fellowship
under award AST-0302123.

\end{acknowledgements}

\end{document}